\documentclass[a4paper, conference]{IEEEtran}
\ifCLASSINFOpdf
\else
\fi
\usepackage{bbm}
\usepackage{amsmath}
\usepackage{amsfonts}
\usepackage{mathrsfs}
\usepackage{amssymb}

 \newtheorem{thm}{Theorem}[section]

 \newtheorem{prop}[thm]{Proposition}
\newtheorem{cor}[thm]{Corollary}
\newtheorem{exm}[thm]{Example}
 \newtheorem{conj}[thm]{Conjecture}

\def\f#1{{\mathbb{F}}_{#1}}

\begin{document}
%
\title{On Deep Holes of Projective Reed-Solomon Codes}

\author{\IEEEauthorblockN{Jun Zhang}
\IEEEauthorblockA{School of Mathematical Sciences\\
 Capital Normal University, Beijing 100048, China\\
Email: junz@cnu.edu.cn}
\and
\IEEEauthorblockN{Daqing Wan}
\IEEEauthorblockA{Department of Mathematics\\
 University of California, Irvine, CA 92697, USA\\
Email: dwan@math.uci.edu}
}

\maketitle

\begin{abstract}
In this paper, we obtain new results on the covering radius and deep holes for projective Reed-Solomon (PRS) codes.
\end{abstract}


%
\IEEEpeerreviewmaketitle

\section{Introduction}
\subsection{Notations and the Main Results}
Let $\f{q}^n$ be the $n$-dimensional vector space over the finite field $\f{q}$ of $q$ elements with characteristic $p$. For any vector (also, called \emph{word}) $ {x}=(x_1,x_2,\cdots,x_n)\in \f{q}^n$, the \emph{Hamming weight} $\mathrm{Wt}( {x})$ of $ {x}$ is defined to be the number of non-zero coordinates, i.e.,
$\mathrm{Wt}( {x})=\#\left\{i\,|\,1\leqslant i\leqslant n,\,x_i\neq 0\right\}.$
For integers $0\leq k\leq n$,
a \emph{linear $[n,k]$ code} $C$ is a $k$-dimensional linear subspace of $\f{q}^n$. The \emph{minimum distance} $d(C)$ of $C$ is the minimum Hamming weight among all non-zero vectors in $C$, i.e.,
$d(C)=\min\{\mathrm{Wt}( {c})\,|\, {c}\in C\setminus\{ {0}\}\}.$
A linear $[n,k]$ code $C\subseteq \f{q}^n$ is called a $[n,k,d]$ linear
code if $C$ has minimum distance $d$. A well-known trade-off between
the parameters of a linear $[n,k,d]$ code is the Singleton bound
which states that
$d\leqslant n-k+1.$
 An $[n,k,d]$ code is called a \emph{maximum distance separable}
  (MDS) code if $d=n-k+1$. An important class of MDS codes are affine Reed-Solomon codes and projective Reed-Solomon codes, which will be our main object
  of study in this paper.

 Let $C$ be an $[n,k,d]$ linear code over $\f{q}$. The \emph{error distance} of any word $u\in\f{q}^n$ to $C$ is defined to be
$$d(u,C)=\min\{d(u,v)\,|\,v\in C\},$$
where $$d(u,v)=\#\{i\,|\,u_{i}\neq v_{i},\,1\le i\le n\}$$
is the Hamming distance between words $u$ and $v$.
The error distance plays an important role in the decoding of the code. The maximum error distance
\[
   \rho(C)=\max\{d(u,\, C)\,|\,u\in \f{q}^n\}
\]
is called the \emph{covering radius} of $C$.

Covering radius of codes was studied extensively~\cite{CKMS85,CLS86,GS85,HKM78,MCL84,OST99}.
For MDS codes, the covering radius is known to be either $d-1$ or $d-2$~\cite{GK98}. For a general
MDS code, determining the exact covering radius is difficult. We shall see below that
affine Reed-Solomon codes have covering radius $d-1$. In contrast, the covering radius of projective Reed-Solomon codes
is unknown in general but is conjectured to be $d-2$. We now recall the definition of Reed-Solomon codes.

Fix a subset
$D=\{x_1,\ldots,x_n\}\subseteq \f{q}$, which is
called the evaluation set. For integer $0<k<n$, the affine RS code
$C=RS(D,k)$ of length $n$ and dimension $k$ over
$\f{q}$ is defined to be
\begin{align*}
    RS(D,k)=\{(f(x_1),\ldots,f(x_n))\in
\f{q}^n\,|\, f(x)\in \f{q}[x],\\ \deg f(x)\leq k-1\}.
\end{align*}

It is easy to check that the minimal distance of this code is $n-k+1$, and thus $RS(D, k)$ is a MDS code.
For $D=\f{q}$, we write $RS(q,k)$ for short.

For any word $u\in \f{q}^{n}$, by the Lagrange interpolation, there is a unique
polynomial $f$ of degree $\le n-1$ such that
\[
   u=u_{f}=(f(x_{1}),f(x_{2}),\cdots,f(x_{n})).
\]
Clearly, $u_{f}\in RS(D,k)$ if and only if $\deg(f)\le k-1$. We also say that $u_{f}$ is defined by the polynomial $f(x)$.
One can easily show~(see \cite{LW08}):
for any $k \leqslant \deg(f)\leqslant n-1$, we have the inequality
 \[
   n-\deg(f)\leqslant d(u_{f},RS(D,k))\leqslant n-k.
\]
It follows that if $\deg(f)=k$, then $d(u_{f},RS(D,k))=n-k$.  One deduces the following

\begin{prop}\label{RSC}
The covering radius of RS codes with parameters $[n,k,d=n-k+1]$ is $n-k=d-1$.
\end{prop}

If the distance from a word to the code achieves the covering radius of the code, then the word is called a \emph{deep hole} of the code. Deciding deep holes of a given code is much harder than the covering radius problem, even for RS codes. The deep hole problem for RS codes was studied in~\cite{CMP11,CM07,KW15,LW08,LZ15,Liao11,WL08,WH12,ZFL13}. 
As noted above, words $u_f$ with $\deg(f)=k$ are deep hole of $RS(D, k)$.
Based on numerical computations, Cheng and Murray~\cite{CM07} conjectured that the converse is also true if $D=\f{q}$.

\begin{conj}[\cite{CM07}]\label{deepholeconj} For $0<k<q$,
a word $u_f$ is a deep hole of $RS(q, k)$ if and only if $\deg(f)=k$.
\end{conj}

This conjecture remains open, but has been proved in \cite{ZCL16}
if either $k+1\le p$ or $3\le q-p+1\le k+1\le q-2$. In particular, the conjecture is true for prime fields.
The aim of this paper is to try to extend the above results and conjecture to projective Reed-Solomon codes.
This turns out to be more difficult, as the covering radius is already unknown. For simplicity, we shall assume that $q$ is odd.

Recall that the Projective Reed-Solomon (PRS) code is defined to be
\begin{align*}
    PRS(q+1,k)=\{(f(\alpha_1),\cdots,f(\alpha_q),c_{k-1}(f))\,|\,f(x)\in \f{q},\\ \deg(f(x))<k\}
\end{align*}
where $\f{q}=\{\alpha_1,\alpha_2,\cdots,\alpha_q=0\}$ and $c_{k-1}(f)$ is the coefficient of the term of degree $k-1$ of $f(x)$. In other words, $PRS(q+1,k)$ has one generator matrix of the form
\begin{equation*}
  \left(
    \begin{array}{ccccc}
      1 & 1 & \cdots & 1 & 0 \\
      \alpha_1 & \alpha_2 & \cdots & \alpha_q & 0 \\
      \vdots & \vdots & \ddots & \vdots & \vdots \\
      \alpha_1^{k-2} & \alpha_2^{k-2} & \cdots & \alpha_q^{k-2} & 0 \\
      \alpha_1^{k-1} & \alpha_2^{k-1} & \cdots & \alpha_q^{k-1} & 1 \\
    \end{array}
  \right).
\end{equation*}
It is easy to check that the PRS code has minimum distance $q+2-k$ and thus it is also an MDS code.

For the case $k=1$, the PRS code $PRS(q+1,1)$ is nothing but the repeating code generated by $(1,1,\cdots,1)$.
In this case, one can easily show that the covering radius is $d-2=q-1$ and the deep holes are
permutations of $\f{q}\cup \{\alpha\}$, where $\alpha\in\f{q}$ is arbitrary.

For the case $k=q-1$, the proof of Theorem~\ref{thm1} in the next section shows that the covering radius of $PRS(q+1,k)$ is $d-2=1$ and the deep holes are $(a,\cdots,a,0,v)+PRS(q+1,k)$, where $a,v\in \f{q}$ are arbitrary with $a\neq 0$.

For the case $k=q$, one can show that the covering radius of $PRS(q+1,k)$ is $d-1=1$ and the deep holes are $w+PRS(q+1,k)$ for all $w\in\f{q}^{q+1}$ of weight $1$.

With the boundary cases removed, we can then assume that  $2\le k\le q-2$.

Although the covering radius of RS codes is always $d-1$, it seems a little
surprising that the covering radius of PRS codes is unknown in general.
The example in~\cite{CKMS85} is one PRS code $C$ over $\f{5}$ with generator matrix
 \begin{equation*}
\left(
  \begin{array}{cccccc}
    1 & 1 & 1 & 1 & 1 & 0 \\
    1 & 2 & 3 & 4 & 0 & 0 \\
    1 & 2^2 & 3^2 & 4^2 & 0 & 0 \\
    1 & 2^3 & 3^3 & 4^3 & 0 & 1 \\
  \end{array}
\right).
\end{equation*}
The code $C$ has minimum distance $3$ and covering radius $1$.
 This example suggests $PRS(q+1,k)$ may have covering radius $q-k$, two smaller than the minimum distance $q+2-k$.
 This leads to

 \begin{conj}[Covering radius for PRS codes]\label{CR} For odd $q$,
the covering radius of the projective Reed-Solomon code $PRS(q+1,k)$ is $d-2=q-k$.
\end{conj}

In~\cite{Arne94}, D\"{u}r proved

\begin{prop}[\cite{Arne94}]\label{old}
Let $q$ be odd. If $2\le k<\frac{\sqrt{q}}{4}+\frac{39}{16}$ or $6\sqrt{q\ln q}-2\le k\le q-2$, then the covering radius of $PRS(q+1,k)$ is
\[
   \rho(PRS(q+1,k))=q-k.
\]
\end{prop}

Our first result is to improve Proposition~\ref{old} in the cases $q=p$ and $q=p^2$. Using recent results of Ball~\cite{Ball12} and Ball-De Beule~\cite{BB12}
on Conjecture~\ref{mds}, we prove

\begin{thm}\label{improvement}
Let $\f{q}$ be a finite field of $q$ elements and of odd characteristic $p$.
\begin{enumerate}
  \item If $q=p$, for any $2\le k\le p-2$, the covering radius of $PRS(p+1,k)$ is
\[
   \rho(PRS(p+1,k))=p-k.
\]
  \item If $q=p^2$, for $2\le k\le 2p-3$, the covering radius of $PRS(q+1,k)$ is
\[
   \rho(PRS(q+1,k))=q-k.
\]
\end{enumerate}
\end{thm}

The first part shows that the covering radius conjecture is true in the case $q=p$, solving the open case ($q=13,\,k=4$) proposed in~\cite{Arne94}.

Our second result is on deep holes of PRS codes. To describe it, we need to introduce more definitions.
As the first $q$ coordinates of $PRS(q+1,k)$ corresponds to an affine part of the projective line, we represent a vector in $\f{q}^{q+1}$ by $(u_f,v)$ where $u_f\in\f{q}^q$ is defined by a polynomial $f$ of degree $\le q-1$:
\[
   u_f=\left(f(\alpha_1),\cdots,f(\alpha_q)\right)
\]
 and $v\in \f{q}$ is arbitrary. It is easy to see that $(u_f,v)\in \f{q}^{q+1}$ represents a codeword of $PRS(q+1,k)$ if and only if
 \[
    \deg(f)\le k-1\,\mbox{and}\,v=c_{k-1}(f).
 \]

\begin{thm}\label{thm1}
Let $q$ be odd. Assume that Conjecture~\ref{deepholeconj} is true. The covering radius of $PRS(q+1,k)$ is $q-k$,  and
the words $\{(u_f,v)\,|\,\deg(f)=k,\,v\in\f{q}\}$ are deep holes of $PRS(q+1,k)$.
\end{thm}

This first part shows that the covering radius conjecture is a consequence of  the conjecture on deep holes of $RS(q,k)$,
providing an additional evidence to the covering radius conjecture. Since Conjecture~\ref{deepholeconj} is known to be true
if $k+1\le p$ or $3\le q-p+1\le k+1\le q-2$, we deduce

\begin{cor}\label{cor}
Let $q$ be odd. Assume $3\le k+1\le p$ or $3\le q-p+1\le k+1\le q-2$. The covering radius of $PRS(q+1,k)$ is $q-k$,  and
the words $\{(u_f,v)\,|\,\deg(f)=k,\,v\in\f{q}\}$ are deep holes of $PRS(q+1,k)$.
\end{cor}

A further question is to classify all deep holes of the PRS codes. In this direction, we propose

\begin{conj}\label{conj1}
Let $q$ be odd. For $2\le k\le q-2$,
the set $\{(u_f,v)\,|\,\deg(f)=k,\,v\in\f{q}\}$ are all the deep holes of $PRS(q+1,k)$.
\end{conj}

This conjecture is stronger than the covering radius conjecture.
As a positive evidence, we prove

\begin{thm}\label{deepholes}
If $\deg{f}\ge k+1$, denote $s=\deg(f)-k+1$. There are positive constants $c_1$ and $c_2$ such that if
$$s<c_1\sqrt{q},\,\left(\frac{s}{2}+2\right)\log_2(q)<k<c_2q,$$
then for any $v\in \f{q}$, $(u_f,v)$ is not a deep hole of $PRS(q+1,k)$.
\end{thm}


{\sl Remark}. From the results on covering radii of RS codes, PRS codes and MDS codes, one may expect that there is some relationship between the covering radius and the minimum distance for general codes. The example below shows that there does not exist such a relationship in general.

\begin{exm}
Let $RS(n,k)$ be an RS code over $\f{q}$. For any word $u\in \f{q}^n\setminus RS(n,k)$, let
$C_u=\f{q}u\oplus RS(n,k).$
Then the minimum distance $d(C_u)$ is
$d(C_u)=d(u, RS(n,k)).$
If we take $u$ to be the vector defined by $x^k$, we obtain $C_u=RS(n,k+1)$. So the covering radius of $C_u=d(C_u)-1=n-k-1$.
However, if we take $u$ to be any $1$-error vector, e.g. $u=(1,0,0,\cdots,0)$, we obtain a code $C_u$ with minimum distance $1$. However, $C_u$ has large covering radius
$$\ge \rho(RS(n,k))-M(RS(n,k),RS(n,k+1))=n-k-1,$$
by the following proposition.
\end{exm}
\begin{prop}[\cite{CKMS85}]
Let $C_1\subset C_2$ and denote
$$M(C_1,\, C_2)=\max\{d(c,C_1)\,|\,c\in C_2\}.$$ Then
\[
   \rho(C_1)\le \rho(C_2)+M(C_1,\, C_2).
\]

\end{prop}

\subsection{Extension: Covering Radius of the longest MDS Codes}

We first recall the MDS conjecture.
\begin{conj}[MDS Conjecture]\label{mds}
For every linear $[n, k]$ MDS code over $\f{q}$, if $1 < k < q$,
then $n\le q+1$, except when $q$ is even and $k=3$ or $k=q-1$, in which cases $n\le q + 2$.
\end{conj}

In~\cite{Arne94}, D\"{u}r proved that the covering radius of $PRS(q+1,k)$ is $q-k$ if and only if the normal rational curve is complete (see~\cite{Sto92} for improvements) in the projective geometry $PG(q-k,\, q)$. From this, he deduced that the covering radius conjecture for PRS codes is a consequence of the MDS conjecture.
The covering radius of MDS codes constructed from elliptic curves was studied in~\cite{OST99}. Recently, authors in~\cite{BGP15} used elliptic curves to construct infinite families of MDS codes with covering radius $d-2$, but of length $<q+1$. Actually, the length of most MDS codes constructed from elliptic curves is automatically $<q+1$ from~\cite{LWZ15,Mun92}.

 \begin{flushleft}
   \textbf{ Assumption:} For simplicity, from now on, we assume that the size $q$ of the finite field $\f{q}$ is odd.
\end{flushleft}

One reason is that deep holes of RS codes over finite fields of even characteristic may be more complex~\cite{ZFL13}. Another reason is that in the odd $q$ case,
$q+1$ corresponds to the longest length of MDS codes over $\f{q}$ according to MDS conjecture.

In this paper, we only consider MDS codes of the longest length derived from MDS conjecture.
Could we say something more about the structure on the longest MDS codes, i.e., $n=q+1$?
From Subsection A, we saw that the PRS code $PRS(q+1,k)$ is an MDS code of length $q+1$. Conversely, do PRS codes of length $q+1$ form all the MDS codes of length $q+1$? This is related to a problem about the structure of $(q+1)$-arc in finite geometry proposed by Segre in~\cite{Segre55}.

\begin{prop}[\cite{Ball12}]\label{ball12}
For $k\le p$ or $3\le q-p+1\le k\le q-2$, the length of MDS codes over $\f{q}$ can not exceed $q+1$. Moreover, for the range of $k$ above, if
the length $n$ of an MDS code $C\,[n, k]$ over $\f{q}$ achieves $q+1$, then $C$ is equivalent to the PRS code $PRS(q+1, k)$.
\end{prop}

By Propositions~\ref{ball12} and the above results on covering radius of PRS codes, we obtain
\begin{thm}\label{thm2}
Let $q$ be odd. For $3\le k+1\le p$ or $3\le q-p+1\le k+1\le q-2$, the covering radius of any MDS code $C$ with parameters $[q+1,k]$ is
\[
   \rho(C)=q-k.
\]
\end{thm}

In particular, we have
\begin{cor}
Let $p$ be a prime. For $2\le k\le p-2$, the covering radius of any MDS code $C$ with parameters $[p+1,k]$ is
\[
   \rho(C)=p-k.
\]
\end{cor}

Similar to Conjecture~\ref{conj1}, we propose
\begin{conj}\label{conj2}[General covering radius conjecture]
Let $q$ be odd. For $2\le k\le q-2$, the covering radius of any MDS code $C$ with parameters $[q+1,k]$ is
\[
   \rho(C)=q-k.
\]
\end{conj}

{\sl Remark.}
If all the MDS codes of length $q+1$ over the finite field $\f{q}$ ($q$ odd) are equivalent to the PRS codes of length $q+1$, then Conjecture~\ref{conj2} immediately follows from Conjecture~\ref{CR}. However, there is an MDS code of length $10$ over $\f{9}$ discovered by Glynn~\cite{Glynn86} which is not equivalent to the PRS code $PRS(10, 5)$. This is the only one MDS code known so far which is not equivalent to PRS codes. This MDS code has a generator matrix
 \begin{equation*}
  \left(
    \begin{array}{ccccc}
      1 & 1 & \cdots & 1 & 0 \\
      \alpha_1 & \alpha_2 & \cdots & \alpha_9 & 0 \\
       \alpha_1^2+w\alpha_1^6 & \alpha_2^2+w\alpha_2^6 & \cdots & \alpha_9^2+w\alpha_9^6 & 0 \\
      \alpha_1^{3} & \alpha_2^{3} & \cdots & \alpha_9^{3} & 0 \\
      \alpha_1^{4} & \alpha_2^{4} & \cdots & \alpha_9^{4} & 1 \\
    \end{array}
  \right),
\end{equation*}
where $\f{9}=\{\alpha_1,\alpha_2,\cdots,\alpha_9\}$ and $w$ is any fixed element in $\f{9}$ such that $w^4+1\neq 0$. Using mathematical softwares, one can check the code has covering radius $4$ which also satisfies Conjecture~\ref{conj2}.

\section{Proofs of Theorems~\ref{improvement},~\ref{thm1} and~\ref{deepholes}}
\subsection{Deep Holes of $RS(q,k)$ and Proof of Theorem~\ref{thm1}}
We first review some results about error distances and deep holes of RS codes $RS(q,k)$ which will help us prove Theorem~\ref{thm1}.

Recall that all deep holes of $RS(q, k)$ were conjectured to be defined by polynomials of degree $k$ (Conjecture~\ref{deepholeconj}).
The same conjecture is false for general evaluation set $D$, see~\cite{WH12,ZFL13}.
To prove Conjecture~\ref{deepholeconj}, the easiest case is to determine whether a polynomial of degree $k+1$ defines a deep hole of $RS(q, k)$. Li and Wan~\cite{LW08} interpreted it as a subset sum problem (SSP). Let $G$ be a finite abelian group and $D$ a subset of $G$. The $k$-SSP over $D$ consists in determining for any $g\in G$, if there is a subset $S\subset D$ such that $|S|=k$ and $\sum_{s\in S}s=g$.
For general $D$, solving $k$-SSP is an \textbf{NP}-hard problem. But in our case, we only care $D=G=\f{q}$ where the $k$-SSP is easy.
\begin{prop}[\cite{LW08}]\label{ssp}
If $D=G=\f{q}$, then for any $g\in \f{q}$ and for any $1\le k\le q-1$, the $k$-SSP over $D$ always has solutions.
\end{prop}
The above proposition is applied to decide deep holes of $RS(q, k)$.
\begin{prop}[\cite{LW08}]
For any $1\le k\le q-2$, the vectors defined by polynomials of degree $k+1$ are not deep holes of $RS(q, k)$.
\end{prop}

For general degrees, the authors in~\cite{CM07} got the first result by reducing the conjecture to the
existence of rational points on a hypersurface over $\f{q}$.
Following a similar approach of Cheng-Wan~\cite{CW07}, Li and Wan~\cite{WL08} improved
the result in~\cite{CM07} with Weil's character sum estimate.
Later, Cafure et.al.~\cite{CMP11} improved the result in~\cite{WL08} a little bit by using tools of algebraic geometry.

Liao~\cite{Liao11} gave a tighter estimation of error distance, which was improved by Zhu and Wan~\cite{ZW12}.
\begin{prop}[\cite{ZW12}]\label{ZW12}
Let $u_f \in\f{q}^q$ such that $1\le r\le d = \deg(f)-k < q-1-k$. There are positive constants $c_1$ and $c_2$ such that if
$$d<c_1\sqrt{q},\,((d+r)/2+1)\log_2(q)<k<c_2q,$$
 then $d(u_f, RS(q, k))\le q-k-r$.
\end{prop}

A major progress in proving Conjecture~\ref{deepholeconj} is the recent result:
\begin{prop}[\cite{ZCL16}]\label{CLZ15}
If $k+1\le p$ or $3\le q-p+1\le k+1\le q-2$, Conjecture~\ref{deepholeconj} is true.
\end{prop}

\begin{flushleft}
\textbf{Proof of Theorem~\ref{thm1}.}
\end{flushleft}
Our task is to compute the maximal error distance $d((u_f,v),\,PRS(q+1,k))$ for all vectors $(u_f,v)\in\f{q}^{q+1}$.
\begin{enumerate}
  \item If $\deg(f)\le k-1$, it is easy to see that $d((u_f,v),\,PRS(q+1,k))\le 1$.
  \item If $\deg(f)=k$, without loss of generality, assume $f(x)=x^k+ax^{k-1}$, since the terms of degree $\le k-2$ could be killed by codewords in $(RS(q,k-1),0)\subset PRS(q+1,k)$. To compute the error distance $d((u_f,v),\,PRS(q+1,k))$, it is equivalent to finding some codeword $(u_g,c_{k-1}(g))\in\f{q}^{q+1}$ maximizing the number of zeros in the error vector $(u_f-u_g,v-c_{k-1}(g))=(u_{f-g},v-c_{k-1}(g))$. On one hand, the number of zeros in the vector $(u_{f-g},v-c_{k-1}(g))$ is $\le k+1$, as $f-g$ which is a polynomial of degree $k$ has at most $k$ zeros. On the other hand, we next show that $k+1$ is achievable. This forces $c_{k-1}(g)=v$ and $f-g$ has $k$ zeros in $\f{q}$. In other words, we need to find a polynomial $h\in\f{q}[x]$ of degree $\le k-2$ such that $f-vx^{k-1}-h=x^k-(v-a)x^{k-1}-h$ has $k$ zeros. This is equivalent to the following $k$-SSP
      \[
        S\subset \f{q},\,\mbox{subject to}\, |S|=k\mbox{ and}\,\sum_{s\in S}s=v-a,
      \]
      having solutions. By Proposition~\ref{ssp}, this $k$-SSP always has solutions. So there exists a polynomial $g$ of degree at most $k-1$ such that $c_{k-1}(g)=v$ and $f-g$ has $k$ zeros in $\f{q}$. In conclusion, if $\deg(f)=k$, then
      \[
       d((u_f,v),\,PRS(q+1,k))=q-k.
      \]
  \item\label{3rdcase} If $\deg(f)\ge k+1$, we have
     \[
    d((u_f,v),\,PRS(q+1,k))\le d(u_f,\,RS(q,k))+1.
     \]
     By Conjecture~\ref{deepholeconj}, we have
     \[
     d(u_f,\,RS(q,k))\le q-k-1.\]
It follows that
     \[
    d((u_f,v),\,PRS(q+1,k))\le q-k-1+1=q-k.
     \]
\end{enumerate}
Putting the above three cases together, we obtain the covering radius of $PRS(q+1,k)$
\[
   \rho=\max\{d((u_f,v),\,PRS(q+1,k))\,|\,(u_f,v)\in\f{q}^{q+1}\}=q-k.
\]
In addition,  for any polynomial $f$ of degree $k$ and any $v\in \f{q}$, the vector $(u_f,v)$ is a deep hole of $PRS(q+1,k)$.

\subsection{Proof of Theorem~\ref{improvement}}
For all vectors $(u_f,v)\in\f{q}^{q+1}$, the error distance
     \[
  d((u_f,v),\,PRS(q+1,k))\le d(u_f,\,RS(q,k))+1\le q-k+1.
     \]
In addition, if $\deg(f)=k$, then
      \[
       d((u_f,v),\,PRS(q+1,k))=q-k.
      \]
So the covering radius of $PRS(q+1,k)$ satisfies
     \[
    q-k\le \rho(PRS(q+1,k))\le q-k+1.
     \]
If $\rho(PRS(q+1,k))= q-k+1$, then there is some vector $w=(w_1,\cdots,w_q,w_{q+1})\in \f{q}^{q+1}$ such that
\[
  d(w,\,PRS(q+1,k))=q-k+1.
\]
In this case, the new code
\[
    C_w=\f{q}w\oplus PRS(q+1,k)
\]
is an MDS code which has a generator matrix
\begin{equation*}
  \left(
    \begin{array}{ccccc}
      1 & 1 & \cdots & 1 & 0 \\
      \alpha_1 & \alpha_2 & \cdots & \alpha_q & 0 \\
      \vdots & \vdots & \ddots & \vdots & \vdots \\
      \alpha_1^{k-1} & \alpha_2^{k-1} & \cdots & \alpha_q^{k-1} & 1 \\
      w_1 & w_2 & \cdots & w_q & w_{q+1} \\
    \end{array}
  \right)
\end{equation*}
where $\f{q}=\{\alpha_1, \alpha_2, \cdots, \alpha_q\}$. So one can easily check that the following matrix
\begin{equation*}
  \left(
    \begin{array}{cccccc}
      1 & 1 & \cdots & 1 & 0 & 0 \\
      \alpha_1 & \alpha_2 & \cdots & \alpha_q & 0 & 0\\
      \vdots & \vdots & \ddots & \vdots & \vdots& \vdots \\
      \alpha_1^{k-1} & \alpha_2^{k-1} & \cdots & \alpha_q^{k-1} & 1 & 0 \\
      w_1 & w_2 & \cdots & w_q & w_{q+1}& 1 \\
    \end{array}
  \right)
\end{equation*}
generates a $[q+2,k+1]$ MDS code.
\begin{enumerate}
  \item By Proposition~\ref{ball12}, when $q=p$ is a prime, for all $k+1\le p$,
there is no MDS code with parameters $[q+2,k+1]$, which contradicts to our assumption $\rho(PRS(q+1,k))= q-k+1$. So
the first part of Theorem~\ref{improvement} is proved.
  \item For general $q=p^a\,(a\ge 2)$, Ball-De Beule~\cite{BB12} improved the result in~\cite{Ball12}: for all $k+1\le 2p-2$, there is no
  MDS code with parameters $[q+2,k+1]$. So we get the second part of Theorem~\ref{improvement}.
\end{enumerate}

\subsection{Discussion: Deep Holes of $PRS(q+1,k)$}

From the proof of Theorem~\ref{thm1} above, we have seen that
 \begin{enumerate}
   \item  polynomials $f$ of degree $<k$ never define any deep hole of $PRS(q+1,k)$;
   \item  polynomials $f$ of degree $k$ define deep holes $(u_f,v)$ of $PRS(q+1,k)$.
 \end{enumerate}
 So we only need to investigate whether
a polynomial $f$ of degree $\ge k+1$ and $v\in \f{q}$ define a deep hole $(u_f,v)$ of $PRS(q+1,k)$.

From the case~\ref{3rdcase}) in Proof of Theorem~\ref{thm1}, a polynomial $f$ of degree $\ge k+1$ and $v\in \f{q}$ define a deep hole $(u_f,v)$ of $PRS(q+1,k)$ if and only if
   \[
     d(u_f,\,RS(q,k))= q-k-1, \mbox{(provided that Conjecture~\ref{deepholeconj} holds)}
   \]
   and for any polynomial $g\in\f{q}[x]$ of degree $\le k-1$ such that $d(u_f,\,u_g)= q-k-1$, we have
$c_{k-1}(g)\neq v.$
So to prove any polynomial $f$ of degree $\ge k+1$ and $v\in \f{q}$ do not define a deep hole, one can follow this idea to construct a polynomial $g\in\f{q}[x]$ of degree $\le k-1$ such that $d(u_f,\,u_g)= q-k-1$, but
$
      c_{k-1}(g)= v.
$

On the contrary, we will directly start from the following inequality:
\begin{equation*}
    \begin{array}{rl}
      &d((u_f,v),\,PRS(q+1,k))\\
       =& \min\{ d((u_f,v),\,(u_g,c_{k-1}(g)))\,|\,\deg(g)\le k-1\} \\
       \le &   \min\{ d((u_f,v),\,(u_g,v))\,|\,\deg(g)= k-1, c_{k-1}(g)=v\}\\
       =&d(u_{f-vx^{k-1}},\,RS(q,k-1)).\\
    \end{array}
\end{equation*}
Even if Conjecture~\ref{deepholeconj} holds, then
\[
 d(u_{f-vx^{k-1}},\,RS(q,k-1))\le q-(k-1)-1=q-k,
\]
but we still can not exclude such a $(u_f,v)$ as a deep hole. Here, we need more accurate information of the error distance $d(u_{f-vx^{k-1}},\,RS(q,k-1))$. Even a little bit more, which ensures $d(u_{f-vx^{k-1}},\,RS(q,k-1))\le q-k-1$, is enough! Taking $r=2$ and $s=\deg(f)-k+1$ in Proposition~\ref{ZW12}, there are positive constants $c_1$ and $c_2$ such that if
$$s<c_1\sqrt{q},\,(\frac{s}{2}+2)\log_2(q)<k<c_2q,$$
 then $d(u_f, RS(q, k-1))\le q-k-1$. So in this case,\
 \begin{align*}
     d((u_f,v),\,PRS(q+1,k))&\le d(u_{f-vx^{k-1}},\,RS(q,k-1))\\
     &\le q-k-1,
 \end{align*}
and hence, we obtain Theorem~\ref{deepholes}.


\section*{Acknowledgement}
This paper was written when the first two authors were visiting Beijing International Center for Mathematical Research (BICMR). The authors would like to thank Prof. Ruochuan Liu for his hospitality. The research of J. Zhang was supported by Beijing outstanding talent training program (No.2014000020124G140). The research of D. Wan was supported by National Science Foundation.

\bibliographystyle{IEEETranS}
\bibliography{covering}
\end{document}